\documentclass[a4paper,12pt]{iopart}
\usepackage{iopams}
\usepackage{setstack}
\usepackage{graphicx}
\usepackage{bm}
\usepackage{epsfig}

\newcommand{\diag}{{\rm diag\,}}

\newcommand{\Str}{{\rm Str\,}}
\newcommand{\Sdet}{{\rm Sdet\,}}
\newcommand{\Pf}{{\rm Pf\,}}

\newcommand{\USp}{{\rm USp\,}}
\newcommand{\U}{{\rm U\,}}

\newcommand{\Ber}{{\rm Ber\,}}
\newcommand{\B}{{\rm B\,}}
\newcommand{\RE}{{\rm Re\,}}

\newcommand{\IM}{{\rm Im\,}}
\newcommand{\Vol}{{\rm Vol\,}}
\newcommand{\eins}{\leavevmode\hbox{\small1\kern-3.8pt\normalsize1}}
\eqnobysec

\begin{document}

\newtheorem{definition}{Definition}[section]
\newtheorem{assumption}[definition]{Assumption}
\newtheorem{theorem}[definition]{Theorem}
\newtheorem{lemma}[definition]{Lemma}
\newtheorem{corollary}[definition]{Corollary}

\title[Pfaffians in $\beta=2$ ensembles]{Surprising Pfaffian factorizations in Random Matrix Theory with Dyson index $\beta=2$}
\author{Mario Kieburg}
\address{Department of Physics and Astronomy, State University of New York at Stony Brook, NY 11794-3800, USA}
\eads{\mailto{mario.kieburg@stonybrook.edu}}

\date{\today}

\begin{abstract}
In the past decades, determinants and Pfaffians were found for eigenvalue correlations of various random matrix ensembles. These structures simplify the average over a large number of ratios of characteristic polynomials to integrations over one and two characteristic polynomials only. Up to now it was thought that determinants occur for ensembles with Dyson index $\beta=2$ whereas Pfaffians only for ensembles with $\beta=1,4$. We derive a non-trivial Pfaffian determinant for $\beta=2$ random matrix ensembles which is similar to the one for $\beta=1,4$. Thus, it unveils a hidden universality of this structure. We also give a general relation between the orthogonal polynomials related to the determinantal structure and the skew-orthogonal polynomials corresponding to the Pfaffian.  As a particular example we 
consider the chiral unitary ensembles in great detail.
\end{abstract}


\submitto{\JPA}

\section{Introduction}\label{sec1}

Random matrix ensembles serve as simple models in a wide range of applications \cite{Efe97,Guhr:1997ve,Verbaarschot:2000dy,Meh04,ABF11} which can be found in number theory \cite{Mont73,BogKea}, disordered systems \cite{Efe97}, quantum chaos \cite{Guhr:1996wn}, empirical data analysis \cite{LCBP99,STRF08,Recher}, information theory \cite{Braunstein:2009gv}, and quantum chromodynamics (QCD) \cite{Shuryak:1992pi}. The complexity of most systems prevents derivations of correlation functions whereas analytic results are accessible for the corresponding random matrix model. The reason for the applicability of random matrix theory lies in the universality of spectral statistics on certain scales like the local mean level spacing \cite{ADMN97,Akemann:2003vy,MZ10} or on the global scale \cite{AJM93,Brezin:1994sq,Ambjorn:1996ga,Akemann:1996zr}. If the Lagrangian of the physical system drastically simplifies such that it is effectively described by global symmetries there might be a random matrix model fulfilling the same symmetries.

Already in the 60's and 70's \cite{Meh60,MehGau60,Dys62,Dys70,Meh71}, the $k$-point correlation functions of the Gaussian and circular ensembles for the three symmetries of orthogonal ($\beta=1$; GOE/COE), unitary ($\beta=2$; GUE/CUE) and unitary-symplectic ($\beta=4$; GSE/CSE) invariance were derived. They can be expressed as a single determinant for the unitary case and a single Pfaffian for $\beta\in\{1,4\}$ where the integrals are pulled inside of these structures. Their matrix elements only depend on two eigenvalues which is a drastic simplification of the integrand. Since then many other random matrix ensembles were studied, e.g. the Ginibre ensembles \cite{Gin65,LehSom91,Ake01,SomWie08} and the the other two rotation groups ${\rm O}(N)$ and $\USp(2N)$ \cite{HPZ05}. The $k$-point correlation functions as well as the averages over ratios of characteristic polynomials for many of these ensembles are determinants and Pfaffians with relatively simple entries only depending on one or two eigenvalues \cite{BreHik00,MehNor01,BorStr05}. For a long time it was thought that determinants appear for ensembles with $\beta=2$ and Pfaffians for the other two cases. In Refs.~\cite{KieGuh09a,KieGuh09b} the general conditions where derived to find these structures. Thus all these particular random matrix ensembles were unified in one procedure to derive these structures.

Very recently a random matrix model for the Wilson Dirac operator was introduced \cite{Damgaard:2010cz} in lattice QCD. It generalizes the chiral GUE which was studied in a Hermitian version \cite{Damgaard:2010cz,Akemann:2010zp,arXiv:1105.6229,arXiv:1108.3035} and a non-Hermitian one \cite{Kieburg:2011uf}. The eigenvalue correlations exhibit Pfaffians for the Hermitian \cite{arXiv:1108.3035} as well as for the non-Hermitian case \cite{KVZ11} reflecting the structure found in Ref.~\cite{KieGuh09b}. This structure has to be also valid in the continuum limit which is the chiral GUE. Hence the question arises if the Pfaffian determinants obtained for the $k$-point correlation functions and thus for the averages over ratios of characteristic polynomials are much more general than conjectured in the broad literature.

Also in other intermediate random matrix ensembles Pfaffians were found. For example a similar situation arises in the transition from GUE to GOE or GSE \cite{Pandey:1982br,KieGuh09a}. If the ensemble is purely a GUE then then the eigenvalue correlations can be cast into determinants whereas the smallest interaction with a GOE or a GSE yields a Pfaffian. It would be of theoretical, technical and numerical interest if all ensembles corresponding to $\beta=2$ exhibit this phenomenon when coupling it to another random matrix ensemble. Such a property simplifies the spectral statistics of intermediate ensembles onto the behavior of the entries of the Pfaffian which are averages of one or two characteristic polynomials only.

Recently, Forrester and Sinclair introduced Pfaffians at $\beta=2$. In Ref.~\cite{Sin} Sinclair extends the Pfaffian found for the partition function with $\beta=1,4$ to Hyperpfaffians with $\beta=L^2,L^2+1$ ($L\in\mathbb{N}$) which also comprises the $\beta=2$ case. With help of these results the authors of Ref.~\cite{ForSin} studied a ${\rm log}$-gas on a ring with two interacting species. One component of this gas is described by a $\beta=4$ ${\rm log}$ gas and the other one by a $\beta=1,2$ ${\rm log}$ gas. The Pfaffian determinants found in Refs.~\cite{Sin,ForSin} are similar to but not the same as the one derived in Sec.~\ref{sec3}.

We derive Pfaffian determinants for averages over ratios of characteristic polynomials weighted by a joint probability density function factorizing in weights of the single eigenvalues apart from a squared Vandermonde determinant. This squared Vandermonde determinant can be cast into one determinant similar to the $\beta=4$ case. Thus it fulfills the same condition as presented in Ref.~\cite{KieGuh09b} which implies a Pfaffian. This unifies all ten symmetry classes in the Cartan classification \cite{Altland:1997zz,Zir96} and exhibits a hidden universal algebraic property in all of these ensembles.

An introduction of the main idea and of the important functions for the technique used here is given in Sec.~\ref{sec0}. In Sec.~\ref{sec2}, we recall some basics known about the determinantal structure obtained for averages over ratios of characteristic polynomials with respect to chiral unitary random matrix ensembles. In contrast to this structure we derive Pfaffians for the same correlation functions in Sec.~\ref{sec3}. Thereby we discuss the Wilson-Dirac random matrix ensemble as a neat application and a good motivation of the derived Pfaffian determinant at the end of this section. The skew-orthogonal polynomials corresponding to the Pfaffian determinants are indeed closely related to the orthogonal polynomials which are found in the determinantal structures. This relation is shown in Sec.~\ref{sec4}. In Sec.~\ref{sec5}, we discuss the generalization of these results for chiral unitary ensembles to other random matrix ensembles like GUE and CUE.

\section{Preliminaries}\label{sec0}

Structures found in supersymmetry are the key ingredient for the technique used in the ensuing sections. These structures allow to derive determinants as well as Pfaffians of averaged ratios of characteristic polynomials and, thus, $k$-point correlation functions for a large class of random matrix ensembles in a direct way. The main idea is to recognize that these structures are a pure algebraic property of the random matrix ensemble and not an analytic one. By an algebraic rearrangement of the integrand one gets  the determinants and Pfaffians without explicitly calculating any integrals. This idea was first proposed in Refs.~\cite{KieGuh09a,KieGuh09b}.

 The requirements to obtain determinants was traced back to a factorization of the probability density of the random matrix ensemble into densities for the single eigenvalues times two Vandermonde determinants (see Ref.~\cite{KieGuh09a}), i.e. the measure for the single eigenvalues has to be
\begin{eqnarray}\label{0.1}
 d\mu(z)=\prod\limits_{j=1}^Ng_1(z_j)d[z_{j}]|\Delta_N(z)|^2
\end{eqnarray}
with the Vandermonde determinant
\begin{eqnarray}\label{0.2}
 \Delta_{N}(z)&=&\prod\limits_{1\leq a<b\leq N}(z_a-z_b)=(-1)^{N(N-1)/2}\det\left[z_{a}^{b-1}\right]_{1\leq a,b\leq N}.
\end{eqnarray}
 The variables $z$ can be complex which correspond to ensembles related to biorthogonal polynomials \cite{Bergere:2003ht}. For Pfaffians this requirement changes to a weight for pairs of eigenvalues and a single Vandermonde determinant \cite{KieGuh09b}, i.e.
\begin{eqnarray}\label{0.3}
 d\mu(z)=\prod\limits_{j=1}^Ng_2(z_{2j-1},z_{2j})d[z_{2j-1}]d[z_{2j}]\Delta_{2N}(z).
\end{eqnarray}
If one of these two conditions are fulfilled then the technique presented in Refs.~\cite{KieGuh09a,KieGuh09b} circumvents the integration theorem by Dyson and Metha \cite{Dys70,Meh71,Meh04,Gho09}. Moreover the approach of Refs.~\cite{KieGuh09a,KieGuh09b} makes an integration theorem unnecessary at the end since it is automatically fulfilled for random matrix ensembles traced back to measures of the form~\eref{0.1} or \eref{0.3}. This can be readily seen by the combination of the determinantal and Pfaffian factorization for averages over ratios of characteristic polynomials \cite{KieGuh09a,KieGuh09b}, the representation of the orthogonal and skew-orthogonal polynomials as averages of the corresponding ensemble \cite{Sze39,Eyn01,Bergere:2003ht,Meh04,Gho09,arXiv:1005.2983} and the expressions of the kernels of the determinants and Pfaffians in orthogonal and skew-orthogonal polynomials~\cite{Meh04,Gho09}. In Sections~\ref{sec2} and \ref{sec3} we derive the $k$-point correlation function without using the integration theorem by Dyson and Metha.

Although, we do not explicitly need supersymmetry, in particular a superspace, some functions are quite useful to write the algebraic expressions of the calculations in a very compact, constructive and intuitive way. These functions have their origin in the theory of supermatrices. For the interested reader, good introductions in supersymmetry are given in Ref.~\cite{Ber87} and in the appendix of Ref.~\cite{VWZ85}. Here we only recall some of these useful algebraic functions and notions.

A diagonal $(p/q)\times(p/q)$ supermatrix $x$ consists of two blocks, $x=\diag(x_1,x_2)$. The $p\times p$ matrix $x_1$ and the $q\times q$ matrix $x_2$ are indeed diagonal, too. The supertrace ``$\Str$'' and the superdeterminant ``$\Sdet$'' of $s$ is then defined by
\begin{eqnarray}\label{0.4}
 \Str x&=&\tr x_1-\tr x_2=\sum_{j=1}^p x_{j1}-\sum_{i=1}^q x_{i2},\\
 \Sdet x&=&\frac{\det x_1}{\det x_2}=\frac{\prod_{j=1}^p x_{j1}}{\prod_{i=1}^q x_{i2}}.\nonumber
\end{eqnarray}
The crucial function of the method used here is
\begin{eqnarray}\label{0.5}
 \B_{p/q}(x)&=&\frac{\Delta_{p}(x_1)\Delta_{q}(x_2)}{\prod\limits_{a,b}(x_{a1}-x_{b2})}\\
 &=&(-1)^{q(q-1)/2+(q+1)p}\det\left[\begin{array}{c} \displaystyle\left\{\frac{1}{x_{a1}-x_{b2}}\right\}\underset{1\leq b\leq q}{\underset{1\leq a\leq p}{\ }} \\ \displaystyle\left\{x_{b2}^{a-1}\right\}\underset{1\leq b\leq q}{\underset{1\leq a\leq q-p}{\ }} \end{array}\right]\nonumber
\end{eqnarray}
for $p\leq q$. It is the square root of a Berezinian,
\begin{eqnarray}\label{0.6}
 \B_{p/q}^2(x)&=&\Ber_{p/q}^{(2)}(x),
\end{eqnarray}
which is the Jacobian in superspace when diagonalizing a Hermitian $(p/q)\times(p/q)$ supermatrix. The notation on the right hand side of Eq.~\eref{0.6} refers to the one used in Refs.~\cite{KieGuh09a,KieGuh09b}.

Everything we need for the method of Refs.~\cite{KieGuh09a,KieGuh09b} are the functions ``$\Sdet$'' and ``$\B$'' embedded in an ordinary space like $\mathbb{R}^{p+q}$ or $\mathbb{C}^{p+q}$. Hence those readers who are not accustomed to supersymmetry may consider these functions as ordinary, rational functions.

\section{Review of chiral unitary random matrices}\label{sec2}

We consider the anti-Hermitian random matrix
\begin{eqnarray}\label{2.1}
 D=\left[\begin{array}{cc} 0 & W \\ -W^\dagger & 0\end{array}\right]
\end{eqnarray}
which is distributed by the density
\begin{eqnarray}\label{2.2}
 P(D)d[D]=\exp[-\alpha \tr V(WW^\dagger)]\prod\limits_{a,b}d~\RE W_{ab}\ d~\IM W_{ab}
\end{eqnarray}
with a non-zero normalization constant. In particular it serves as a model for the Dirac operator in QCD \cite{Shuryak:1992pi}. The constant $\alpha$ is proportional to $n$. The matrix $W$ is a $n\times(n+\nu)$ rectangular matrix. Each of the $n(n+\nu)$ entries of $W$ is a complex number which might be statistically coupled by the arbitrary density $P$. The parameter $\nu$ with $0\leq\nu\leq n$ is the topological charge or also known as index of the Dirac operator such that $D$ has $\nu$ generic zero eigenmodes. The potential $V$ is invariant under the group $\U(n)$, i.e.
\begin{eqnarray}\label{2.2b}
 V(UWW^\dagger U^\dagger)=UV(WW^\dagger) U^\dagger,
\end{eqnarray}
 and is chosen such that all moments of the ensemble over $\mathbb{C}^{n\times(n+\nu)}$ exist. In the simplest case $P$ is Gaussian. Nevertheless the arguments given here are also true for an arbitrary potential. We only need the property
\begin{eqnarray}\label{2.3}
 P\left(\left[\begin{array}{c|cc} 0 & \Lambda & 0 \\ \hline -\Lambda & 0 & 0 \\ 0 & 0 & 0 \end{array}\right]\right)=\exp[-\alpha \tr V(\Lambda^2)]=\prod\limits_{j=1}^n\exp[-\alpha V(\lambda_j^2)]
\end{eqnarray}
for the matrix $\Lambda=\diag(\lambda_1,\ldots,\lambda_n)$ with the singular values $0\leq\lambda_1\leq\ldots\leq\lambda_n$ of $W$, i.e. there are $U\in\U(n)$ and $V\in\U(n+\nu)$ with
\begin{eqnarray}\label{2.4}
D=\diag(U,V)\left[\begin{array}{c|cc} 0 & \Lambda & 0 \\ \hline -\Lambda & 0 & 0 \\ 0 & 0 & 0 \end{array}\right]\diag(U^\dagger,V^\dagger).
\end{eqnarray}
In this basis the measure~\eref{2.2} can be written as
\begin{eqnarray}\label{2.5}
  P(D)d[D]&=&\frac{\Vol_n\Vol_{n+\nu}}{\Vol_1^n\Vol_\nu}\Delta_{n}^2(\Lambda^2)\prod\limits_{j=1}^n\exp[-\alpha V(\lambda_j^2)]\lambda_j^{2\nu+1}d\lambda_j\\
 &\times&d\mu_{\U(n)/\U^n(1)}(U)d\mu_{\U(n+\nu)/\U(\nu)}(V).\nonumber
\end{eqnarray}
The abbreviation of the constant
\begin{eqnarray}\label{2.7}
 \Vol_l=\prod\limits_{j=1}^l\frac{2\pi^j}{(j-1)!}
\end{eqnarray}
refers to the volume of the unitary group $\U(l)$. Thus, the prefactor in Eq.~\eref{2.5} is the volume of the coset $[\U(n)\times\U(n+\nu)]/[\U^n(1)\times\U(\nu)]$. The measure $d\mu_{\mathfrak{G}}$ is the normalized Haar measure of the coset $\mathfrak{G}$.

An important quantity to analyze the eigenvalue statistics of this ensemble is the average over ratios of characteristic polynomials with respect to $D$, i.e.
\begin{eqnarray}\label{2.8}
 Z_{k_1/k_2}^{(n,\nu)}(\kappa)=\int\limits_{\mathbb{C}^{n\times(n+\nu)}}\frac{\prod\limits_{j=1}^{k_2}\det(D-\imath\kappa_{j2}\eins_{2n+\nu})}{\prod\limits_{j=1}^{k_1}\det(D-\imath\kappa_{j1}\eins_{2n+\nu})}P(D)d[D]
\end{eqnarray}
with the diagonal, non-degenerate $(k_1/k_2)\times(k_1/k_2)$ supermatrix $\kappa=\diag(\kappa_1,\kappa_2)=\diag(\kappa_{11},\ldots,\kappa_{k_11},$ $\kappa_{12},\ldots,\kappa_{k_22})$ and the $2n+\nu$ dimensional unit matrix $\eins_{2n+\nu}$. This average is also known as the partition function with $k_1$ bosonic and $k_2$ fermionic flavors in QCD \cite{Splittorff:2002eb,Fyodorov:2002wq,Akemann:2003vy}. The variables $\kappa_{j1}$ are complex numbers with a non-vanishing imaginary part such that the integral is well defined. The partition function~\eref{2.8} is simply related to the matrix Green function and, thus, to the $k$-point correlation function by derivatives with respect to $\kappa$.

The joint probability density~\eref{2.5} is of the class studied in Ref.~\cite{KieGuh09a} and can, therefore, be written as a determinant. This was derived in many articles before~\cite{Fyodorov:2002md,Splittorff:2002eb,Fyodorov:2002wq}. The crucial idea presented in Ref.~\cite{KieGuh09a} is the combination of the ratio of characteristic polynomials~\eref{2.8} with the two Vandermonde determinants~\eref{2.5} to square roots of Berezinians~\eref{0.5}, i.e.
\begin{eqnarray}\label{2.9}
 \Delta_n^2(\Lambda^2)\frac{\prod\limits_{j=1}^{k_2}\det(\Lambda^2-\kappa_{j2}^2\eins_{n})}{\prod\limits_{j=1}^{k_1}\det(\Lambda^2-\kappa_{j1}^2\eins_{n})}&=&\frac{\B_{l_{11}/l_{21}+n}(\widetilde{\kappa}_1^2,\Lambda^2)\B_{l_{12}/l_{22}+n}(\widetilde{\kappa}_2^2,\Lambda^2)}{\B_{l_{11}/l_{21}}(\widetilde{\kappa}_1^2)\B_{l_{12}/l_{22}}(\widetilde{\kappa}_1^2)}
\end{eqnarray}
for any choice of natural numbers $l_{11}+l_{12}=k_{1}$ and $l_{21}+l_{22}=k_{2}$. 

In Eq.~\eref{2.9}, we split the supermatrix $\kappa$ into the two sets $\widetilde{\kappa}_1=\diag(\widetilde{\kappa}_{11},\widetilde{\kappa}_{21})=\diag(\kappa_{11},\ldots,\kappa_{l_{11}1},\kappa_{12},\ldots,\kappa_{l_{21}2})$ and $\widetilde{\kappa}_2=\diag(\widetilde{\kappa}_{12},\widetilde{\kappa}_{22})=\diag(\kappa_{l_{11}+1,1},\ldots,\kappa_{k_{1}1},\kappa_{l_{21}+1,2},\ldots,\kappa_{k_{2}2})$. The choice how we split this set is arbitrary and, thus, we get equivalent but not trivially related results. This was already recognized by the authors of Ref.~\cite{Akemann:2002vy} for products of characteristic polynomials. Let $d_1=n+l_{21}-l_{11}$ and $d_2=n+l_{22}-l_{12}$. The interesting case is $d_1,d_2\geq0$ because we want to discuss the limit $n\to\infty$ and $k_1$, $k_2$ fixed, at the end of this section. The other cases are discussed in Ref.~\cite{KieGuh09a}.

Without loss of generality we assume $d_1\leq d_2$. We rearrange the integrand~\eref{2.8} with the help of Eq.~\eref{2.9} which yields
\begin{eqnarray}
 \fl Z_{k_1/k_2}^{(n,\nu)}(\kappa)&\propto&\Sdet^{-\nu}\kappa\int\frac{\B_{l_{11}/l_{21}+n}(\widetilde{\kappa}_1^2,\Lambda^2)\B_{l_{12}/l_{22}+n}(\widetilde{\kappa}_2^2,\Lambda^2)}{\B_{l_{11}/l_{21}}(\widetilde{\kappa}_1^2)\B_{l_{12}/l_{22}}(\widetilde{\kappa}_1^2)}\prod\limits_{j=1}^n\exp[-\alpha V(\lambda_j^2)]\lambda_j^{2\nu+1}d\lambda_j\nonumber\\
 \fl&\propto&\Sdet^{-\nu}\kappa\int\frac{\prod\limits_{j=1}^n\exp[-\alpha V(\lambda_j^2)]\lambda_j^{2\nu+1}d\lambda_j}{\B_{l_{11}/l_{21}}(\widetilde{\kappa}_1^2)\B_{l_{12}/l_{22}}(\widetilde{\kappa}_1^2)}\nonumber\\
 \fl&\times&\det\left[\begin{array}{cc} \displaystyle\left\{\frac{1}{\kappa^2_{a1}-\kappa^2_{b2}}\right\}\underset{1\leq b\leq l_{21}}{\underset{1\leq a\leq l_{11}}{\ }} & \displaystyle\left\{\frac{1}{\kappa^2_{a1}-\lambda^2_{b2}}\right\}\underset{1\leq b\leq n}{\underset{1\leq a\leq l_{11}}{\ }} \\ \displaystyle\left\{\kappa^{2(a-1)}_{b2}\right\}\underset{1\leq b\leq l_{21}}{\underset{1\leq a\leq d_1}{\ }} & \displaystyle\left\{\lambda^{2(a-1)}_{b2}\right\}\underset{1\leq b\leq n}{\underset{1\leq a\leq d_1}{\ }} \end{array}\right]\nonumber\\
 \fl&\times&\det\left[\begin{array}{cc} \displaystyle\left\{\frac{1}{\kappa^2_{a1}-\kappa^2_{b2}}\right\}\underset{l_{21}+1\leq b\leq k_2}{\underset{ l_{11}+1\leq a\leq k_1}{\ }} & \displaystyle\left\{\frac{1}{\kappa^2_{a1}-\lambda^2_{b2}}\right\}\underset{1\leq b\leq n}{\underset{ l_{11}+1\leq a\leq k_1}{\ }} \\ \displaystyle\left\{\kappa^{2(a-1)}_{b2}\right\}\underset{l_{21}+1\leq b\leq k_{2}}{\underset{1\leq a\leq d_2}{\ }} & \displaystyle\left\{\lambda^{2(a-1)}_{b2}\right\}\underset{1\leq b\leq n}{\underset{1\leq a\leq d_2}{\ }} \end{array}\right].\label{2.10a}
\end{eqnarray}
Applying the generalized Andr{\'e}ief integration theorem \cite{And1883,KieGuh09a} we obtain
\begin{eqnarray}\label{2.10b}
 \fl Z_{k_1/k_2}^{(n,\nu)}(\kappa)&\propto&\frac{\Sdet^{-\nu}\kappa}{\B_{l_{11}/l_{21}}(\widetilde{\kappa}_1^2)\B_{l_{12}/l_{22}}(\widetilde{\kappa}_1^2)}\\
 \fl&\times&\det\left[\begin{array}{ccc} 0 & \displaystyle\left\{\frac{1}{\kappa^2_{b1}-\kappa^2_{a2}}\right\}\underset{l_{11}+1\leq b\leq k_1}{\underset{ l_{21}+1\leq a\leq k_2}{\ }} & \displaystyle\left\{\kappa^{2(b-1)}_{a2}\right\}\underset{1\leq b\leq d_{2}}{\underset{l_{21}+1\leq a\leq k_2}{\ }} \\ \displaystyle\left\{\frac{1}{\kappa^2_{a1}-\kappa^2_{b2}}\right\}\underset{1\leq b\leq l_{21}}{\underset{1\leq a\leq l_{11}}{\ }} & \displaystyle\left\{F(\kappa_{a1},\kappa_{b1})\right\}\underset{l_{11}+1\leq b\leq k_{1}}{\underset{1\leq a\leq l_{11}}{\ }} & \displaystyle\left\{F_b(\kappa_{a1})\right\}\underset{1\leq b\leq d_{2}}{\underset{1\leq a\leq l_{11}}{\ }} \\ \displaystyle\left\{\kappa^{2(a-1)}_{b2}\right\}\underset{1\leq b\leq l_{21}}{\underset{1\leq a\leq d_1}{\ }} & \displaystyle\left\{F_a(\kappa_{b1})\right\}\underset{l_{11}+1\leq b\leq k_{1}}{\underset{1\leq a\leq d_1}{\ }} & \displaystyle\left\{M_{ab}\right\}\underset{1\leq b\leq d_2}{\underset{1\leq a\leq d_1}{\ }} \end{array}\right]\nonumber
\end{eqnarray}
Notice that Andr{\'e}ief's integration theorem as well as its generalization is only an algebraic rearrangement of the integrals without explicitly calculating any integral. The functions $F$ and $F_a$ are one dimensional integrals and their explicit expressions are not so important as we will see in the discussion after Eq.~\eref{2.10f}. For the interested reader we refer to Ref.~\cite{KieGuh09a} where the explicit integrals are given for general random matrix ensembles corresponding to determinants ($\beta=2$). The constant $d_1\times d_2$ matrix $M=[M_{ab}]$ is given by
\begin{eqnarray}\label{2.10c}
 M_{ab}=\int_{\mathbb{R}} \lambda^{2(a+b-2)}\exp[-\alpha V(\lambda^2)]\lambda^{2\nu+1}d\lambda
\end{eqnarray}
and thus generates the moments of the measure.

In the next step we use the identity
\begin{eqnarray}\label{2.10d}
 \det\left[\begin{array}{cc} A & B \\ C & D \end{array}\right]=\det D\det[ A-BD^{-1}C]
\end{eqnarray}
for arbitrary matrices $A$, $B$ and $C$ and an invertible matrix $D$. For the matrix $D$ we choose the $d_1\times d_1$ matrix
\begin{eqnarray}\label{2.10e}
 D=\displaystyle[M_{ab}]\underset{1\leq a,b\leq d_1}{\ }
\end{eqnarray}
which is only a part of the full rectangular matrix $M$ appearing in Eq.~\eref{2.10b}. The determinant of $D$ is proportional to the normalization constant of the ensemble~\eref{3.2} and $M$ is therefore invertible. Employing Eq.~\eref{2.10d} we find
\begin{eqnarray}\label{2.10f}
 \fl Z_{k_1/k_2}^{(n,\nu)}(\kappa)&=&\frac{1}{\B_{l_{11}/l_{21}}(\widetilde{\kappa}_1^2)\B_{l_{12}/l_{22}}(\widetilde{\kappa}_2^2)}\\
 \fl&\times&\det\left[\begin{array}{cc} \displaystyle\left\{G_1^{(d_1)}(\kappa_{a2},\kappa_{b2})\right\}\underset{l_{21}+1\leq b\leq k_2}{\underset{1\leq a\leq l_{21}}{\ }} & \displaystyle\left\{G_2^{(d_1)}(\kappa_{b1},\kappa_{a2})\right\}\underset{1\leq b\leq l_{11}}{\underset{1\leq a\leq l_{21}}{\ }} \\ \displaystyle\left\{G_2^{(d_1)}(\kappa_{a1},\kappa_{b2})\right\}\underset{l_{21}+1\leq b\leq k_2}{\underset{l_{11}+1\leq a\leq k_1}{\ }} & \displaystyle\left\{G_3^{(d_1)}(\kappa_{a1},\kappa_{b1})\right\}\underset{1\leq b\leq l_{11}}{\underset{l_{11}+1\leq a\leq k_1}{\ }} \\  \displaystyle\left\{H_1^{(a)}(\kappa_{b2})\right\}\underset{l_{21}+1\leq b\leq k_2}{\underset{d_1+1\leq a\leq d_2}{\ }} & \displaystyle\left\{H_2^{(a)}(\kappa_{b1})\right\}\underset{1\leq b\leq l_{11}}{\underset{d_1+1\leq a\leq d_2}{\ }} \end{array}\right]\nonumber.
\end{eqnarray}
In the last step we identify the functions $G_1^{(d_1)}$, $G_2^{(d_1)}$, $G_3^{(d_1)}$, $H_1^{(a)}$ and $H_2^{(a)}$ by considering the particular choices $(l_{11},l_{12},l_{21},l_{22})\in\{(0,0,1,1),(1,0,1,0),(1,1,0,0),(0,0,0,1),$ $(1,0,0,0)\}$. In all of these cases the determinant reduces to one of the entries. Then we obtain
\begin{eqnarray}\label{2.11}
 \fl &&\frac{Z_{k_1/k_2}^{(n,\nu)}(\kappa)}{Z_{0/0}^{(n,\nu)}}=\frac{(-1)^{k_1(k_1-1)/2+(l_{21}+1)(k_1+1)+(l_{11}+1)(k_2+1)}}{\B_{l_{11}/l_{21}}(\widetilde{\kappa}_1^2)\B_{l_{12}/l_{22}}(\widetilde{\kappa}_2^2)}\frac{\prod\limits_{j=0}^{d_1-1}h_j^{(\nu)}}{\prod\limits_{j=0}^{n-1}h_j^{(\nu)}}\\
 \fl&\times&\det\left[\begin{array}{cc} \displaystyle\left\{-\frac{Z_{0/2}^{(d_1-1,\nu)}(\kappa_{a2},\kappa_{b2})}{h_{d_1-1}^{(\nu)}Z_{0/0}^{(d_1-1,\nu)}}\right\}\underset{l_{21}+1\leq b\leq k_2}{\underset{1\leq a\leq l_{21}}{\ }} & \displaystyle\left\{\frac{1}{Z_{0/0}^{(d_1,\nu)}}\frac{Z_{1/1}^{(d_1,\nu)}(\kappa_{b1},\kappa_{a2})}{(\kappa_{b1}^2-\kappa_{a2}^2)}\right\}\underset{1\leq b\leq l_{11}}{\underset{1\leq a\leq l_{21}}{\ }} \\ \displaystyle\left\{\frac{1}{Z_{0/0}^{(d_1,\nu)}}\frac{Z_{1/1}^{(d_1,\nu)}(\kappa_{a1},\kappa_{b2})}{(\kappa_{a1}^2-\kappa_{b2}^2)}\right\}\underset{l_{21}+1\leq b\leq k_2}{\underset{l_{11}+1\leq a\leq k_1}{\ }} & \displaystyle\left\{\frac{h_{d_1}^{(\nu)}}{Z_{0/0}^{(d_1+1,\nu)}}Z_{2/0}^{(d_1+1,\nu)}(\kappa_{a1},\kappa_{b1})\right\}\underset{1\leq b\leq l_{11}}{\underset{l_{11}+1\leq a\leq k_1}{\ }} \\  \displaystyle\left\{\frac{Z_{0/1}^{(a-1,\nu)}(\kappa_{b2})}{Z_{0/0}^{(a-1,\nu)}}\right\}\underset{l_{21}+1\leq b\leq k_2}{\underset{d_1+1\leq a\leq d_2}{\ }} & \displaystyle\left\{\frac{h_{a-1}^{(\nu)}}{Z_{0/0}^{(a,\nu)}}Z_{1/0}^{(a,\nu)}(\kappa_{b1})\right\}\underset{1\leq b\leq l_{11}}{\underset{d_1+1\leq a\leq d_2}{\ }} \end{array}\right]\nonumber
\end{eqnarray}
for the partition function~\eref{2.8} which is a particular result of the general one derived in Ref.~\cite{KieGuh09a}.

The determinant~\eref{2.11} interpolates between one-point and two-point kernels as the entries of the determinant. We emphasize again the choice of the numbers $0\leq l_{11}\leq k_1$ and $0\leq l_{21}\leq k_1$ and the splitting of $\kappa$ are arbitrary. The particular choice $l_{11}=k_1$ and $l_{21}=0$ yields the $k_1+k_2$ dimensional determinant with one-point kernels considered in Refs.~\cite{Splittorff:2002eb,Akemann:2003vy}. This choice is suitable for the microscopic limit in chiral random matrix theory. For bulk and soft edge correlations \cite{Akemann:2003vy} the representation in two point correlations are the better choice to make contact with other random matrix ensembles \cite{AJM93,Brezin:1994sq,Ambjorn:1996ga,MZ10}. This case relates to the choice $l_{11}=k_1$ and $l_{21}=k_2$ for $k_2\leq k_1$ and $l_{11}=0$ and $l_{21}=0$ for $k_2\geq k_1$.

The $k$-point correlation function at the $k$ variables $x=\diag(x_1,\ldots,x_k)$ is given by
\begin{eqnarray}
 \fl R_k^{(n,\nu)}(x)&\propto&\int_{\mathbb{R}_+^{n-k}}\Delta_n^2(\diag(x^2,\Lambda^2))\exp[-\alpha\tr V(x^2)-\alpha\tr V(\Lambda^2)]{\det}^{2\nu+1}x\prod\limits_{j=1}^{n-k}\lambda_j^{2\nu+1}d\lambda_j\nonumber\\
 \fl&\propto&\Delta_k^2(x^2){\det}\, x\exp[-\alpha\tr V(x^2)]Z_{0/2k}^{(n-k,\nu)}(\diag(x,-x)).\label{2.12a}
\end{eqnarray}
Now we employ the formula~\eref{2.11} for $(l_{11},l_{12},l_{21},l_{22})=(0,0,k,k)$ and find the result
\begin{eqnarray}
 \fl R_k^{(n,\nu)}(x)&\propto&\det\left[\sqrt{x_ax_b}\exp[-\alpha( V(x_a^2)+V(x_b^2))/2]Z_{0/2}^{(n-1,\nu)}(x_a,-x_b)\right]_{1\leq a,b\leq k}.\label{2.12b}
\end{eqnarray}
Since $Z_{0/2}^{(n-1,\nu)}(x_a,-x_b)=(-1)^\nu Z_{0/2}^{(n-1,\nu)}(x_a,x_b)$ this agrees with the general formula for $\beta=2$ ensembles \cite{Meh04}. Please notice that we derived this formula without using the integration theorem by Dyson and Mehta \cite{Dys70,Meh71,Meh04,Gho09}.

The constant $h_j^{(\nu)}$ in Eq.~\eref{2.11} is the normalization constant of the orthogonal polynomial
\begin{equation}\label{2.12}
 p_j^{(\nu)}(x^2)=\frac{(-1)^{j}}{(-\imath x)^\nu}\frac{Z_{0/1}^{(j,\nu)}(x)}{Z_{0/0}^{(j,\nu)}}.
\end{equation}
These polynomials solve the orthogonality relation
\begin{equation}\label{2.13}
 \int\limits_{0}^\infty p_j^{(\nu)}(x^2)p_i^{(\nu)}(x^2)x^{2\nu+1}\exp[-\alpha V(x^2)]dx=h_j^{(\nu)}\delta_{ji}.
\end{equation}
The authors of Ref.~\cite{Damgaard:1997ye} have shown that these polynomials fulfill a recursion relation with respect to the topological charge $\nu$ by
\begin{equation}\label{2.13b}
 \frac{p_j^{(\nu+1)}(x)}{p_j^{(\nu+1)}(0)}=\frac{1}{x}\frac{p_j^{(\nu)}(0)p_{j+1}^{(\nu)}(x)-p_{j+1}^{(\nu)}(0)p_j^{(\nu)}(x)}{p_j^{(\nu)}(0)p_{j+1}^{(\nu)\prime}(0)-p_{j+1}^{(\nu)}(0)p_j^{(\nu)\prime}(0)}
\end{equation}
which is quite useful by taking the limit $n\to\infty$. This relation follows when setting $m=0$ in Eq.~(12) of Ref.~\cite{Damgaard:1997ye}. One can readily prove identity~\eref{2.13b} by showing the orthogonality relation~\eref{2.13} for the right hand side with respect to the $\nu+1$ measure, i.e.
\begin{eqnarray}
 \fl&&\int\limits_{0}^\infty \frac{p_j^{(\nu)}(0)p_{j+1}^{(\nu)}(x^2)-p_{j+1}^{(\nu)}(0)p_j^{(\nu)}(x^2)}{x^2}\frac{p_l^{(\nu)}(0)p_{l+1}^{(\nu)}(x^2)-p_{l+1}^{(\nu)}(0)p_l^{(\nu)}(x^2)}{x^2}x^{2\nu+3}e^{-\alpha V(x^2)}dx\nonumber\\
 \fl&\propto&\int\limits_{0}^\infty \sum_{a=0}^{j}\frac{p_a^{(\nu)}(0)p_a^{(\nu)}(x^2)}{h_a^{(\nu)}}(p_l^{(\nu)}(0)p_{l+1}^{(\nu)}(x^2)-p_{l+1}^{(\nu)}(0)p_l^{(\nu)}(x^2))x^{2\nu+1}e^{-\alpha V(x^2)}dx\nonumber\\
 \fl&\propto&\delta_{jl}\label{2.13c},
\end{eqnarray}
where we used the Christoffel-Darboux formula. The monic normalization of $p_j^{(\nu)}(x)=x^j+\ldots$ for all $j$ and $\nu$ explains the choice of the constants.

The Cauchy transform of $p_j^{(\nu)}$ is related to the partition function with one bosonic flavor by
\begin{eqnarray}\label{2.14}
 \widehat{p}_{j}^{(\nu)}(x^2)&=&\int\limits_{0}^\infty\frac{p_j^{(\nu)}(\lambda^2)}{\lambda^2-x^2}\lambda^{2\nu+1}\exp[-\alpha V(\lambda^2)]d\lambda\\
 &=&(-1)^{j}(-\imath x)^\nu\frac{h_{j}^{(\nu)}}{Z_{0/0}^{(j+1,\nu)}}Z_{1/0}^{(j+1,\nu)}(x)\nonumber.
\end{eqnarray}
In the result~\eref{2.11} we recognize that the choices $(l_{11},l_{12},l_{21},l_{22})=(0,0,1,1),(1,0,1,0),$ $(1,1,0,0)$ yield the same partition functions as the choices $(l_{11},l_{12},l_{21},l_{22})=(0,0,0,2),(1,0,0,1),(2,0,0,0)$, respectively. Therefore the two-flavor partition functions in Eq.~\eref{2.11} can also be expressed in the orthogonal polynomials~\eref{2.12} and their Cauchy transforms~\eref{2.14}, i.e.
\begin{eqnarray}
 \fl\frac{Z_{0/2}^{(d_1-1,\nu)}(\kappa_{a2},\kappa_{b2})}{Z_{0/0}^{(d_1-1,\nu)}}&=&-\frac{(-\kappa_{a2}\kappa_{b2})^\nu}{\kappa_{a2}^2-\kappa_{b2}^2}\det\left[\begin{array}{cc} p_{d_1-1}^{(\nu)}(\kappa_{a2}^2) & p_{d_1-1}^{(\nu)}(\kappa_{b2}^2) \\ p_{d_1}^{(\nu)}(\kappa_{a2}^2) & p_{d_1}^{(\nu)}(\kappa_{b2}^2) \end{array}\right]\label{2.15},\\
 \fl\frac{Z_{2/0}^{(d_1+1,\nu)}(\kappa_{a1},\kappa_{b1})}{Z_{0/0}^{(d_1+1,\nu)}}&=&\frac{1}{h_{d_1}^{(\nu)}h_{d_1-1}^{(\nu)}}\frac{1}{(-\kappa_{a1}\kappa_{b1})^\nu(\kappa_{a1}^2-\kappa_{b1}^2)}\det\left[\begin{array}{cc} \widehat{p}_{d_1-1}^{(\nu)}(\kappa_{a1}^2) & \widehat{p}_{d_1-1}^{(\nu)}(\kappa_{b1}^2) \\ \widehat{p}_{d_1}^{(\nu)}(\kappa_{a1}^2) & \widehat{p}_{d_1}^{(\nu)}(\kappa_{b1}^2) \end{array}\right],\nonumber\\
 &&\label{2.16}\\
 \fl\frac{Z_{1/1}^{(d_1,\nu)}(\kappa_{a1},\kappa_{b2})}{Z_{0/0}^{(d_1,\nu)}}&=&\frac{1}{h_{d_1-1}^{(\nu)}}\left(\frac{\kappa_{b2}}{\kappa_{a1}}\right)^\nu\det\left[\begin{array}{cc} \widehat{p}_{d_1-1}^{(\nu)}(\kappa_{a1}^2) & p_{d_1-1}^{(\nu)}(\kappa_{b2}^2) \\ \widehat{p}_{d_1}^{(\nu)}(\kappa_{a1}^2) & p_{d_1}^{(\nu)}(\kappa_{b2}^2) \end{array}\right]\label{2.17}.
\end{eqnarray}
These three relations are already well known \cite{Meh04,Gho09}. They can also be derived with help of the Christoffel-Darboux formula.

The structure~\eref{2.11} is a general property of ensembles with a joint probability density including a squared Vandermonde determinant as considered in Sec.~4.2 of Ref.~\cite{KieGuh09a} whereas the relations~\eref{2.15}-\eref{2.17} have to be slightly modified for other ensembles.

In the microscopic limit the authors of Refs.~\cite{ADMN97,Akemann:2003vy} have shown that for a generic potential $V$ the orthogonal polynomials and their Cauchy transforms become
\begin{eqnarray}
 p_n^{(\nu)}\left(\frac{x^2}{(cn)^2}\right)&\overset{n\gg1}{\propto}&\frac{J_{\nu}(x)}{x^{\nu}},\label{2.18}\\
 \widehat{p}_n^{(\nu)}\left(\frac{x^2}{(cn)^2}\right)&\overset{n\gg1}{\propto}&x^{\nu}K_{\nu}(x),\label{2.19}
\end{eqnarray}
where $c$ is a constant depending on the potential $V$. The functions $J_{\nu}$ and $K_{\nu}$ are the Bessel function of the first kind and the modified one of the second kind, respectively. Hence in the microscopic limit the partition function~\eref{2.8} is
\begin{eqnarray}\label{2.20}
 \fl Z_{k_1/k_2}^{(n,\nu)}\left(\frac{\kappa}{cn}\right)&\overset{n\gg1}{\propto}&\frac{1}{\B_{l_{11}/l_{21}}(\widetilde{\kappa}_1^2)\B_{l_{12}/l_{22}}(\widetilde{\kappa}_2^2)}\\
 \fl&&\times\det\left[\begin{array}{cc} \displaystyle\left\{ I^{(1)}_\nu(\kappa_{a2},\kappa_{b2})\right\}\underset{l_{21}+1\leq b\leq k_2}{\underset{1\leq a\leq l_{21}}{\ }} & \displaystyle\left\{ I^{(2)}_\nu(\kappa_{b1},\kappa_{a2})\right\}\underset{1\leq b\leq l_{11}}{\underset{1\leq a\leq l_{21}}{\ }} \\ \displaystyle\left\{ I^{(2)}_\nu(\kappa_{a1},\kappa_{b2})\right\}\underset{l_{21}+1\leq b\leq k_2}{\underset{l_{11}+1\leq a\leq k_1}{\ }} & \displaystyle\left\{ I^{(3)}_\nu(\kappa_{a1},\kappa_{b1})\right\}\underset{1\leq b\leq l_{11}}{\underset{l_{11}+1\leq a\leq k_1}{\ }} \\ \displaystyle\left\{\kappa_{b2}^{a}J_{\nu+a}(\kappa_{b2})\right\}\underset{l_{21}+1\leq b\leq k_2}{\underset{0\leq a\leq d_2-d_1-1}{\ }} & \displaystyle\left\{\kappa_{b1}^{a}K_{\nu+a}(\kappa_{b1})\right\}\underset{1\leq b\leq l_{11}}{\underset{0\leq a\leq d_2-d_1-1}{\ }} \end{array}\right]\nonumber,
\end{eqnarray}
where
\begin{eqnarray}
 \fl I^{(1)}_\nu(\kappa_{a2},\kappa_{b2})&=&\left\{\begin{array}{cl} \displaystyle\frac{\kappa_{a2}J_{\nu-1}(\kappa_{a2})J_{\nu}(\kappa_{b2})-\kappa_{b2}J_{\nu}(\kappa_{a2})J_{\nu-1}(\kappa_{b2})}{\kappa_{a2}^2-\kappa_{b2}^2}, & a\neq b, \\ \displaystyle\frac{J_{\nu+1}(\kappa_{a2})J_{\nu-1}(\kappa_{a2})-J_{\nu}^2(\kappa_{a2})}{2}, & a= b, \end{array}\right.\label{2.21}\\
 \fl I^{(2)}_\nu(\kappa_{a1},\kappa_{b2})&=&\frac{\kappa_{a1}K_{\nu-1}(\kappa_{a1})J_{\nu}(\kappa_{b2})-\kappa_{b2}K_{\nu}(\kappa_{a1})J_{\nu-1}(\kappa_{b2})}{\kappa_{a1}^2-\kappa_{b2}^2},\label{2.22}\\
 \fl I^{(3)}_\nu(\kappa_{a1},\kappa_{b1})&=&\left\{\begin{array}{cl} \displaystyle\frac{\kappa_{a1}K_{\nu-1}(\kappa_{a1})K_{\nu}(\kappa_{b1})-\kappa_{b1}K_{\nu}(\kappa_{a1})K_{\nu-1}(\kappa_{b1})}{\kappa_{a1}^2-\kappa_{b1}^2}, & a\neq b, \\ \displaystyle\frac{K_{\nu+1}(\kappa_{a1})K_{\nu-1}(\kappa_{a1})-K_{\nu}^2(\kappa_{a1})}{2}, & a= b. \end{array}\right.\label{2.23}
\end{eqnarray}
This is the well known result found in the literature~\cite{Fyodorov:2002md,Splittorff:2002eb,Fyodorov:2002wq}.

\section{Derivation of the Pfaffian determinant}\label{sec3}

In subsection~\ref{sec3.1} we derive a Pfaffian determinant for the same class of chiral random matrix ensembles discussed in Sec.~\ref{sec2}. A neat application of this Pfaffian is presented in subsection~\ref{sec3.2}. This example is the random matrix model for the Wilson-Dirac operator in lattice QCD \cite{Damgaard:2010cz,Akemann:2010zp,arXiv:1105.6229,arXiv:1108.3035,Kieburg:2011uf}.

\subsection{Pfaffian determinants in chiral random matrix theory}\label{sec3.1}

We show that the representations in determinants~\eref{2.11} are not the only existing ones for chiral unitary ensembles. A non-trivial Pfaffian can be derived for the partition function by noticing that the square of the Vandermonde in the measure~\eref{2.5} can be rewritten as one Vandermonde determinant of the variables $\pm\lambda_j$, i.e.
\begin{equation}\label{3.1}
 \Delta_{n}^2(\Lambda^2)=(-1)^{n(n-1)/2}\frac{\Delta_{2n}(\Lambda,-\Lambda)}{2^n\det\Lambda}.
\end{equation}
The determinant of $\Lambda$ will be put into the weight later on, cf. Eqs.~\eref{3.2} and \eref{3.3} below. Considering the Wilson random matrix theory \cite{Damgaard:2010cz,Akemann:2010zp,arXiv:1105.6229,arXiv:1108.3035,Kieburg:2011uf} such a splitting arises in a natural way for finite lattice spacing. Then an eigenvalue pair $\pm\imath\lambda_j$ becomes either a complex conjugated pair or two independent real eigenvalues corresponding to a pair of eigenvectors with positive and negative chirality. Hence, the Pfaffian resulting from the single Vandermonde determinant~\eref{3.1} is the one which is generalized to non-zero lattice spacing and not the determinant \cite{arXiv:1108.3035,KVZ11}.

This allows us to define an anti-symmetric two-point measure on $\mathbb{R}^2$
\begin{equation}\label{3.2}
 \fl g(x_1,x_2)=\frac{|x_1x_2|^\nu}{4}\exp\left[-\alpha \frac{V(x_1^2)+V(x_2^2)}{2}\right]\delta(x_1+x_2)[\Theta(x_1)-\Theta(x_2)],
\end{equation}
where $\Theta$ is the Heaviside distribution. Then we consider the measure
\begin{equation}\label{3.3}
 D[\lambda]=\frac{\Vol_n\Vol_{n+\nu}}{\Vol_1^n\Vol_\nu}\Delta_{2n}(\lambda)\prod\limits_{j=1}^ng(\lambda_{2j-1},\lambda_{2j})d\lambda_{2j}d\lambda_{2j-1}
\end{equation}
over $2n$ independent eigenvalues instead of the measure~\eref{2.5}. This measure fulfills the general condition for finding a Pfaffian, cf. Ref.~\cite{KieGuh09b} and see also Eq.~\eref{0.3}.

The partition function~\eref{2.8} can be expressed in terms of this new measure,
\begin{eqnarray}\label{3.4}
 Z_{k_1/k_2}^{(n,\nu)}(\kappa)=\frac{(-1)^{n(k_1+k_2)}}{n!}\Sdet^{-\nu}(-\imath\kappa)\int\limits_{\mathbb{R}^{2n}}\prod\limits_{a=1}^{2n}\frac{\prod\limits_{j=1}^{k_2}(\kappa_{j2}-\lambda_a)}{\prod\limits_{j=1}^{k_1}(\kappa_{j1}-\lambda_a)}D[\lambda].
\end{eqnarray}
In the first step we extend the Vandermonde determinant~\eref{3.1} with the characteristic polynomials,
\begin{eqnarray}\label{3.5}
  Z_{k_1/k_2}^{(n,\nu)}(\kappa)&=&(-1)^{n(k_1+k_2)}\frac{\Vol_n\Vol_{n+\nu}}{n!\Vol_1^n\Vol_\nu}\Sdet^{-\nu}(-\imath\kappa)\\
&\times&\int\limits_{\mathbb{R}^{2n}}\frac{\B_{k_1/k_2+2n}(\kappa,\lambda)}{\B_{k_1/k_2}(\kappa)}\prod\limits_{j=1}^ng(\lambda_{2j},\lambda_{2j-1})d\lambda_{2j}d\lambda_{2j-1}.\nonumber
\end{eqnarray}
This representation is apart from the $z_{2N+1}$-integral of the form as in Eq.~(3.3) in Ref.~\cite{KieGuh09b}. Notice that in this extension we do not have the same freedom as in the determinantal case~\eref{2.9} since there is only one Vandermonde determinant in the integrand~\eref{3.3}. Let $d=2n+k_2-k_1\geq0$. Then we employ the representation of the function ``$\B$'' as a determinant, see Eq.~\eref{0.5},
\begin{eqnarray}\label{3.5a}
  Z_{k_1/k_2}^{(n,\nu)}(\kappa)&\propto&\frac{\Sdet^{-\nu}\kappa}{\B_{k_1/k_2}(\kappa)}\int\limits_{\mathbb{R}^{2n}}\prod\limits_{j=1}^ng(\lambda_{2j},\lambda_{2j-1})d\lambda_{2j}d\lambda_{2j-1}\\
&\times&\det\left[\begin{array}{cc} \displaystyle\left\{\frac{1}{\kappa_{a1}-\kappa_{b2}}\right\}\underset{1\leq b\leq k_2}{\underset{ 1\leq a\leq k_1}{\ }} & \displaystyle\left\{\frac{1}{\kappa_{a1}-\lambda_{b2}}\right\}\underset{1\leq b\leq 2n}{\underset{ 1\leq a\leq k_1}{\ }} \\ \displaystyle\left\{\kappa^{a-1}_{b2}\right\}\underset{1\leq b\leq k_{2}}{\underset{1\leq a\leq d}{\ }} & \displaystyle\left\{\lambda^{a-1}_{b2}\right\}\underset{1\leq b\leq 2n}{\underset{1\leq a\leq d}{\ }} \end{array}\right]\nonumber.
\end{eqnarray}
The generalized de Bruijn integration theorem \cite{Bru55,KieGuh09a} can be applied now which yields
\begin{eqnarray}\label{3.5b}
 \fl Z_{k_1/k_2}^{(n,\nu)}(\kappa)&\propto&\frac{\Sdet^{-\nu}\kappa}{\B_{k_1/k_2}(\kappa)}\\
 \fl&\times&\Pf\left[\begin{array}{ccc} 0 & \displaystyle\left\{\frac{1}{\kappa_{b1}-\kappa_{a2}}\right\}\underset{1\leq b\leq k_1}{\underset{ 1\leq a\leq k_2}{\ }} & \displaystyle\left\{\kappa^{b-1}_{a2}\right\}\underset{1\leq b\leq d}{\underset{1\leq a\leq k_2}{\ }} \\ \displaystyle\left\{-\frac{1}{\kappa_{a1}-\kappa_{b2}}\right\}\underset{1\leq b\leq k_2}{\underset{ 1\leq a\leq k_1}{\ }} & \displaystyle\left\{\widetilde{F}(\kappa_{a1},\kappa_{b1})\right\}\underset{1\leq a,b\leq k_1}{\ } & \displaystyle\left\{\widetilde{F}_b(\kappa_{a1})\right\}\underset{1\leq b\leq d}{\underset{1\leq a\leq k_1}{\ }} \\ \displaystyle\left\{-\kappa^{a-1}_{b2}\right\}\underset{1\leq b\leq k_{2}}{\underset{1\leq a\leq d}{\ }} & \displaystyle\left\{-\widetilde{F}_a(\kappa_{b1})\right\}\underset{1\leq b\leq k_1}{\underset{1\leq a\leq d}{\ }} & \displaystyle\left\{\widetilde{M}_{ab}\right\}\underset{1\leq a,b\leq d}{\ } \end{array}\right].\nonumber
\end{eqnarray}
As Andr{\'e}ief's integration theorem the generalized de Bruijn integration theorem is only an algebraic rearrangement of the integrals without calculating any of them. The functions $\widetilde{F}$ and $\widetilde{F}_a$ are two-fold integrals and are again not much of importance, see the discussion after Eq.~\eref{3.5e}. Explicit expressions of them are given in Ref.~\cite{KieGuh09b} for general random matrix ensembles corresponding to Pfaffians comprising the measure~\eref{3.3}, too.

The $d\times d$ anti-symmetric matrix $\widetilde{M}=[\widetilde{M}_{ab}]$ consists of the moments
\begin{eqnarray}\label{3.5c}
 \widetilde{M}_{ab}=\int_{\mathbb{R}^2}(\lambda_1^{a-1}\lambda_2^{b-1}-\lambda_1^{b-1}\lambda_2^{a-1})g(\lambda_1,\lambda_2)d\lambda_1d\lambda_2.
\end{eqnarray}
Analogously to Eq.~\eref{2.10d}, we employ the identity
\begin{eqnarray}\label{3.5d}
 \Pf\left[\begin{array}{cc} A & B \\ -B^T & C \end{array}\right]=\Pf C\, \Pf[A+BC^{-1}B^T]
\end{eqnarray}
with an arbitrary matrix $B$, an arbitrary antisymmetric matrix $A$ and an arbitrary even dimensional, antisymmetric matrix $C$ which has to be invertible. Let $k_1+k_2$ be even. Then $d$ is also even and the Pfaffian of the matrix $\widetilde{M}$ is proportional to the normalization constant of the ensemble~\eref{2.2}. Hence the choice $C=\widetilde{M}$ is well-defined. This yields 
\begin{eqnarray}\label{3.5e}
  \fl Z_{k_1/k_2}^{(n,\nu)}(\kappa)&\propto&\frac{1}{\B_{k_1/k_2}(\kappa)}\\
 \fl&\times&\Pf\left[\begin{array}{cc}  \displaystyle\left\{\widetilde{G}_1^{(d)}(\kappa_{a2},\kappa_{b2})\right\}\underset{1\leq a,b\leq k_2}{\ } & \displaystyle\left\{\widetilde{G}_2^{(d)}(\kappa_{b1},\kappa_{a2})\right\}\underset{1\leq b\leq k_1}{\underset{ 1\leq a\leq k_2}{\ }}  \\ \displaystyle\left\{-\widetilde{G}_2^{(d)}(\kappa_{a1},\kappa_{b2})\right\}\underset{1\leq b\leq k_2}{\underset{ 1\leq a\leq k_1}{\ }} & \displaystyle\left\{\widetilde{G}_3^{(d)}(\kappa_{a1},\kappa_{b1})\right\}\underset{1\leq a,b\leq k_1}{\ } \end{array}\right]\nonumber.
\end{eqnarray}
The functions $\widetilde{G}_1^{(d)}$, $\widetilde{G}_2^{(d)}$ and $\widetilde{G}_3^{(d)}$ can be obtained by considering the cases $(k_1/k_2)=(0/2),(1/1),(2/0)$, respectively. In each of these cases the Pfaffian~\eref{3.5e} reduces to a single term.
This leads to a particular case of the general result derived in Ref.~\cite{KieGuh09b}. We find our main result of this article
\begin{eqnarray}\label{3.6}
  \fl&&\frac{Z_{k_1/k_2}^{(n,\nu)}(\kappa)}{Z_{0/0}^{(n,\nu)}}=\frac{(-1)^{k_2(k_2+1)/2}}{\B_{k_1/k_2}(\kappa)}\frac{\prod\limits_{j=0}^{d-1}h_j^{(\nu)}}{\prod\limits_{j=0}^{n-1}h_j^{(\nu)}}\\
 \fl&\times&\Pf\left[\begin{array}{c|c} \displaystyle\underset{}{\frac{\kappa_{b2}-\kappa_{a2}}{h_{d/2-1}^{(\nu)}Z_{0/0}^{(d/2-1,\nu)}}Z_{0/2}^{(d/2-1,\nu)}(\kappa_{a2},\kappa_{b2})} & \displaystyle\frac{1}{Z_{0/0}^{(d/2,\nu)}}\frac{Z_{1/1}^{(d/2,\nu)}(\kappa_{b1},\kappa_{a2})}{(\kappa_{a2}-\kappa_{b1})} \\ \hline \displaystyle\frac{1}{Z_{0/0}^{(d/2,\nu)}}\overset{}{\frac{Z_{1/1}^{(d/2,\nu)}(\kappa_{a1},\kappa_{b2})}{(\kappa_{a1}-\kappa_{b2})}} & \displaystyle\frac{h_{d/2}^{(\nu)}(\kappa_{a1}-\kappa_{b1})}{Z_{0/0}^{(d/2+1,\nu)}}Z_{2/0}^{(d/2+1,\nu)}(\kappa_{a1},\kappa_{b1}) \end{array}\right]\nonumber
\end{eqnarray}
for even $k_1+k_2$. The indices $a$ and $b$ run from $1$ to $k_2$ in the first columns and the first rows and from $1$ to $k_1$ in the last ones. The result for odd $k_2+k_1$ can be readily obtained by introducing an additional fermionic flavor and sending it to infinity. This shifts the parameter $d$ to $d+1$ and adds a row and a column to the matrix in the Pfaffian~\eref{3.6} with the partition functions $Z_{0/1}^{((d-1)/2,\nu)}(\kappa_{b2})$ and $Z_{1/0}^{((d+1)/2,\nu)}(\kappa_{b1})$ which are apart from a factor $\kappa^\nu$ an orthogonal polynomial and its Cauchy-transform, cf. Eqs.~\eref{2.12} and \eref{2.14}. Notice that the matrix in the Pfaffian~\eref{3.6} is indeed antisymmetric because $Z_{0/2}^{(d/2-1,\nu)}$ and $Z_{2/0}^{(d/2+1,\nu)}$ are symmetric under a permutation of the entries.

Indeed, Eq.~\eref{3.6} cannot be traced back to the identity
\begin{equation}\label{3.7}
 \Pf\left[\begin{array}{cc} 0 & X \\ -X^T & 0 \end{array}\right]=(-1)^{p(p-1)/2}\det X
\end{equation}
with an arbitrary $p\times p$ matrix $X$. We refer to the relation~\eref{3.7} as a trivial Pfaffian extension of a determinant. The Pfaffian~\eref{3.6} seems to be the result of recursion relations of the orthogonal polynomials~\eref{2.12}. It is difficult to see how these recursions have to be performed to map the Pfaffian~\eref{3.6} to the determinant~\eref{2.11}. However the construction of this structure seems to be the same for a broad class of ensembles. This is confirmed by the fact that the result~\eref{3.6} can be extended to all factorizing ensembles with a squared Vandermonde determinant in the joint probability density~\eref{0.1}. This will be shown in Sec.~\ref{sec5}.

Again one can consider the $k$-point correlation function~\eref{2.12a} and what it looks like with the Pfaffian determinant. Using the result~\eref{3.6} we find for the $k$-point correlation function
\begin{eqnarray}
 \fl R_k^{(n,\nu)}(x)&\propto&\exp[-\alpha\tr V(x)]\nonumber\\
 \fl&\times&\Pf\left[\begin{array}{c|c} (x_a-x_b)Z_{0/2}^{(n-1,\nu)}(x_a,x_b) & (x_a+x_b)Z_{0/2}^{(n-1,\nu)}(x_a,-x_b) \\ \hline-(x_a+x_b)Z_{0/2}^{(n-1,\nu)}(-x_a,x_b) & -(x_a-x_b)Z_{0/2}^{(n-1,\nu)}(-x_a,-x_b) \end{array}\right]_{1\leq a,b\leq k}\nonumber\\
 \fl&\propto&\exp[-\alpha\tr V(x)]\nonumber\\
 \fl&\times&\Pf\left[\begin{array}{c|c} (x_a-x_b)Z_{0/2}^{(n-1,\nu)}(x_a,x_b) & (x_a+x_b)Z_{0/2}^{(n-1,\nu)}(x_a,x_b) \\ \hline-(x_a+x_b)Z_{0/2}^{(n-1,\nu)}(x_a,x_b) & -(x_a-x_b)Z_{0/2}^{(n-1,\nu)}(x_a,x_b) \end{array}\right]_{1\leq a,b\leq k}.\label{3.7a}
\end{eqnarray}
Again we have not employed the integration theorem by Dyson and Mehta \cite{Dys70,Meh71,Meh04,Gho09}. To see that Eq.~\eref{3.7a} indeed agrees with the determinant~\eref{2.12b} one can consider the square of the Pfaffian,
\begin{eqnarray}
 \fl&&\Pf^2\left[\begin{array}{c|c} (x_a-x_b)Z_{0/2}^{(n-1,\nu)}(x_a,x_b) & (x_a+x_b)Z_{0/2}^{(n-1,\nu)}(x_a,x_b) \\ \hline-(x_a+x_b)Z_{0/2}^{(n-1,\nu)}(x_a,x_b) & -(x_a-x_b)Z_{0/2}^{(n-1,\nu)}(x_a,x_b) \end{array}\right]_{1\leq a,b\leq k}\nonumber\\
 \fl&=&\det\left[\begin{array}{c|c} (x_a-x_b)Z_{0/2}^{(n-1,\nu)}(x_a,x_b) & (x_a+x_b)Z_{0/2}^{(n-1,\nu)}(x_a,x_b) \\ \hline-(x_a+x_b)Z_{0/2}^{(n-1,\nu)}(x_a,x_b) & -(x_a-x_b)Z_{0/2}^{(n-1,\nu)}(x_a,x_b) \end{array}\right]_{1\leq a,b\leq k}\nonumber\\
 \fl&=&2^k\det\left[\begin{array}{c|c} -x_bZ_{0/2}^{(n-1,\nu)}(x_a,x_b) & x_bZ_{0/2}^{(n-1,\nu)}(x_a,x_b) \\ \hline-(x_a+x_b)Z_{0/2}^{(n-1,\nu)}(x_a,x_b) & -(x_a-x_b)Z_{0/2}^{(n-1,\nu)}(x_a,x_b) \end{array}\right]_{1\leq a,b\leq k}\nonumber\\
 \fl&=&2^{2k}\det\left[\begin{array}{c|c} -x_bZ_{0/2}^{(n-1,\nu)}(x_a,x_b) & 0 \\ \hline-(x_a+x_b)Z_{0/2}^{(n-1,\nu)}(x_a,x_b) & -x_aZ_{0/2}^{(n-1,\nu)}(x_a,x_b) \end{array}\right]_{1\leq a,b\leq k}\nonumber\\
 \fl&=&2^{2k}{\det}^2x\,{\det}^2\left[Z_{0/2}^{(n-1,\nu)}(x_a,x_b)\right]_{1\leq a,b\leq k}.\label{3.7b}
\end{eqnarray}
The square root of Eq.~\eref{3.7b} yields Eq.~\eref{2.12b}.

In the large $n$ limit, we employ Eqs.~(\ref{2.15}-\ref{2.19}) and (\ref{2.21}-\ref{2.23}) and obtain
\begin{eqnarray}\label{3.8}
  \fl Z_{k_1/k_2}^{(n,\nu)}\left(\frac{\kappa}{cn}\right)&\overset{n\gg1}{\propto}&\frac{1}{\B_{k_1/k_2}(\kappa)}\\
 \fl&\times&\Pf\left[\begin{array}{c|c} \displaystyle\underset{}{(\kappa_{a2}-\kappa_{b2})I^{(1)}_\nu(\kappa_{a2},\kappa_{b2})} & \displaystyle (\kappa_{b1}+\kappa_{a2})I^{(2)}_\nu(\kappa_{b1},\kappa_{a2}) \\ \hline \displaystyle\overset{}{-(\kappa_{a1}+\kappa_{b2})I^{(2)}_\nu(\kappa_{a1},\kappa_{b2})} & \displaystyle(\kappa_{a1}-\kappa_{b1})I^{(3)}_\nu(\kappa_{a1},\kappa_{b1}) \end{array}\right]\nonumber
\end{eqnarray}
for even $k_1+k_2$ and
\begin{eqnarray}\label{3.9}
  \fl Z_{k_1/k_2}^{(n,\nu)}\left(\frac{\kappa}{cn}\right)&\overset{n\gg1}{\propto}&\frac{1}{\B_{k_1/k_2}(\kappa)}\\
 \fl&\times&\Pf\left[\begin{array}{c|c|c} 0 & \displaystyle\underset{}{J_\nu(\kappa_{b2})} & \displaystyle K_\nu(\kappa_{b1})
\\ \hline -J_\nu(\kappa_{a2}) & \displaystyle\overset{}{\underset{}{(\kappa_{a2}-\kappa_{b2})I^{(1)}_\nu(\kappa_{a2},\kappa_{b2})}} & \displaystyle (\kappa_{b1}+\kappa_{a2})I^{(2)}_\nu(\kappa_{b1},\kappa_{a2}) \\ \hline -K_\nu(\kappa_{a1}) &  \displaystyle\overset{}{-(\kappa_{a1}+\kappa_{b2})I^{(2)}_\nu(\kappa_{a1},\kappa_{b2})} & \displaystyle(\kappa_{a1}-\kappa_{b1})I^{(3)}_\nu(\kappa_{a1},\kappa_{b1}) \end{array}\right]\nonumber
\end{eqnarray}
for odd $k_1+k_2$. These Pfaffians carry over to the Wilson Dirac random matrix model~\cite{arXiv:1108.3035,KVZ11}. For small numbers of bosonic and fermionic flavors these results were checked by the recursion relations of the Bessel functions \cite{Ver11}.

Please notice the difference in the prefactor of Eqs.~\eref{2.20}, \eref{3.8} and \eref{3.9}. The entries of the Berezinian are the squares of the variables $\kappa$ for the determinantal structure~\eref{2.20} whereas it is only $\kappa$ for the Pfaffian. This yields a technical advantage when calculating eigenvalue correlations of the random matrix models for the Wilson Dirac operator.

\subsection{An application: Wilson-Dirac random matrix}\label{sec3.2}

The Wilson-Dirac operator is a modified Dirac operator on a lattice. In the infrared limit this operator can be modeled by the Wilson-Dirac random matrix \cite{Damgaard:2010cz,Akemann:2010zp,arXiv:1105.6229,Kieburg:2011uf} which is a $(2n+\nu)\times(2n+\nu)$ Hermitian matrix
\begin{eqnarray}\label{3.2.1}
 D_{\rm W}=\left[\begin{array}{cc} a A & W \\ -W^\dagger & a B \end{array}\right]
\end{eqnarray}
distributed by the Gaussian
\begin{eqnarray}\label{3.2.2}
 P(D_{\rm W})=\exp\left[-\frac{n}{2}(\tr A^2+\tr B^2)-n\tr WW^\dagger\right].
\end{eqnarray}
The variable $a$ plays the role of the lattice spacing. The chiral symmetry is explicitly broken by the Hermitian matrices $A$ and $B$, i.e.
\begin{eqnarray}\label{3.2.3}
 \gamma_5\left.D_{\rm W}\right|_{m=0}\gamma_5\neq-\left.D_{\rm W}\right|_{m=0}\quad\mathrm{with}\quad\gamma_5=\diag(\eins_{n},-\eins_{n+\nu}),
\end{eqnarray} 
which have the dimensions $n\times n$ and $(n+\nu)\times(n+\nu)$, respectively.
Hence, $A$ and $B$ model the Wilson-term.

We consider the partition function with $N_f$ fermionic flavors,
\begin{eqnarray}\label{3.2.4}
 Z_{N_{\rm f}}^{(n,\nu)}(m,a)=\int \prod\limits_{j=1}^{N_{\rm f}}\det(D_{\rm W}+m_j\eins_{2n+\nu})P(D_{\rm W})d[D_{\rm W}].
\end{eqnarray}
The external variables $m=\diag(m_1,\ldots,m_{N_{\rm f}})$ play the role of the quark masses. Indeed one can also consider bosonic flavors. However, we restrict ourself to fermionic flavors to keep the example as simple as possible 

In the microscopic limit ($n\to\infty$), $\widehat{m}=2nm$, $\widehat{a}=\sqrt{n}a/2$ and $\nu$ are kept fixed. This yields the integral 
\begin{eqnarray}
 \fl Z_{N_{\rm f}}^{(n,\nu)}(\widehat{m},\widehat{a})&\overset{n\gg1}{=}&\int\limits_{\U(N_f)}\exp\left[\frac{1}{2}\tr\widehat{m}(U+U^{-1})-\widehat{a}^2\tr\left(U^2+U^{-2}\right)\right]{\det}^\nu Ud\mu(U).\label{3.2.5}
\end{eqnarray}
For a derivation of this result we refer to Refs.~\cite{Damgaard:2010cz,Akemann:2010zp}. Exactly the integral~\eref{3.2.5} makes contact with lattice QCD \cite{ShaSin98,RS02,Sha06,necco}.

At zero lattice spacing ($\widehat{a}=0$) this partition function can be identified with the one considered in Sec.~\ref{sec2},
\begin{eqnarray}\label{3.2.6}
 Z_{N_{\rm f}}^{(n,\nu)}(m,a=0)\propto Z_{0/N_{\rm f}}^{(n,\nu)}(\imath m).
\end{eqnarray}
Considering again the microscopic limit~\eref{3.2.5}, we trace the integral back to the $a=0$ result by introducing a $N_{\rm f}\times N_{\rm f}$ Hermitian random matrix $\sigma$ similar to the calculation in Ref.~\cite{arXiv:1105.6229,VerSpl11},
\begin{eqnarray}
 \fl Z_{N_{\rm f}}^{(n,\nu)}(\widehat{m},\widehat{a})&\overset{n\gg1}{\propto}&\int\exp\left[-\frac{1}{4\widehat{a}^2}\tr(\sigma-\imath \widehat{m})^2-2(\widehat{a}N_{\rm f})^2\right]\nonumber\\
 \fl&\times&\int\limits_{\U(N_f)}\exp\left[-\imath\tr\sigma(U+U^{-1})\right]{\det}^\nu Ud\mu(U)d[\sigma]\nonumber\\
 \fl&\propto&\int\exp\left[-\frac{1}{4\widehat{a}^2}\tr(\sigma+\imath \widehat{m})^2-2(\widehat{a}N_{\rm f})^2\right]Z_{0/N_{\rm f}}^{(n,\nu)}(\sigma)d[\sigma].\label{3.2.7}
\end{eqnarray}
Notice that $\sigma$ is an ordinary matrix and not a supermatrix because we consider fermionic flavors, only. The constant $\exp[-2(\widehat{a}N_{\rm f})^2]$ can be shifted into the normalization constant and can, thus, be omitted in the ensuing calculations.

A diagonalization of $\sigma=VsV^\dagger$ with $V\in\U(N_{\rm f})$ yields a Harish-Chandra-Itzykson-Zuber-integral \cite{Har58,ItzZub80} in the Gaussian term. The partition function $Z_{0/N_{\rm f}}^{(n,\nu)}$ is invariant under $\U(N_{\rm f})$. We find
\begin{eqnarray}
 \fl Z_{N_{\rm f}}^{(n,\nu)}(\widehat{m},\widehat{a})&\propto&\int\frac{\det\left[\exp\left[-(s_j-\imath \widehat{m}_i)^2/4\widehat{a}^2\right]\right]_{1\leq j,i\leq N_{\rm f}}}{\Delta_{N_{\rm f}}(\widehat{m})}Z_{0/N_{\rm f}}^{(n,\nu)}(s)\Delta_{N_{\rm f}}(s)d[s].\label{3.2.8}
\end{eqnarray}
Employing the result as a determinant of the microscopic limit of $\widehat{a}=0$ partition function, cf. Eq.~\eref{2.20}, we end up with a complicated expression,
\begin{eqnarray}
 Z_{N_{\rm f}}^{(n,\nu)}(\widehat{m},\widehat{a})&\overset{n\gg1}{\propto}&\int\frac{\det\left[\exp\left[-(s_j-\imath \widehat{m}_i)^2/4\widehat{a}^2\right]\right]_{1\leq j,i\leq N_{\rm f}}}{\Delta_{N_{\rm f}}(\widehat{m})}\nonumber\\
 &\times&\det\left[s_i^{j-1}J_{\nu-1+j}(s)\right]_{1\leq j,i\leq N_{\rm f}}\frac{\Delta_{N_{\rm f}}(s)}{\Delta_{N_{\rm f}}(s^2)}d[s].\label{3.2.9}
\end{eqnarray}
There is no obvious way to further simplify the integral~\eref{3.2.9} due to the factor $\Delta_{N_{\rm f}}(s)/\Delta_{N_{\rm f}}(s^2)$. This was not much of a problem for the authors of Refs.~\cite{arXiv:1105.6229,VerSpl11} because they only considered a small numbers of flavors. However the problem is highly non-trivial for an arbitrary number of flavors.

This problem can be solved by using the Pfaffian expressions~\eref{3.8} and~\eref{3.9} instead of the determinant. Let $N_{\rm f}$ be even to keep the expressions as simple as possible. Then we have for the microscopic limit~\eref{3.2.5}
\begin{eqnarray}
 \fl Z_{N_{\rm f}}^{(n,\nu)}(\widehat{m},\widehat{a})&\overset{n\gg1}{\propto}&\int\frac{\det\left[\exp\left[-(s_j-\imath \widehat{m}_i)^2/4\widehat{a}^2\right]\right]_{1\leq j,i\leq N_{\rm f}}}{\Delta_{N_{\rm f}}(\widehat{m})}\label{3.2.10}\\
 \fl&\times&\Pf\left[\frac{s_jJ_{\nu-1}(s_j)J_\nu(s_i)-s_iJ_{\nu}(s_j)J_{\nu-1}(s_i)}{s_j+s_i}\right]_{1\leq j,i\leq N_{\rm f}}d[s].\nonumber
\end{eqnarray}
After expanding the determinant no term hinders us to pull the integrals into the Pfaffian. We obtain the compact result
\begin{eqnarray}\label{3.2.11}
 \fl Z_{N_{\rm f}}^{(n,\nu)}(\widehat{m},\widehat{a})&\propto&\frac{1}{\Delta_{N_{\rm f}}(\widehat{m})}\Pf\left[(\widehat{m}_j-\widehat{m}_i)Z_{2}^{(n,\nu)}(\widehat{m}_j,\widehat{m}_i,\widehat{a})\right]_{1\leq j,i\leq N_{\rm f}}\nonumber
\end{eqnarray}
with
\begin{eqnarray}\label{3.2.12}
 \fl Z_{2}^{(n,\nu)}(\widehat{m}_1,\widehat{m}_2,\widehat{a})&\propto&\frac{1}{\widehat{m}_1-\widehat{m}_2}\int_{\mathbb{R}^2}\exp\left[-\frac{(s_1-\imath \widehat{m}_1)^2+(s_2-\imath \widehat{m}_2)^2}{4\widehat{a}^2}\right]\\
 \fl&\times&\frac{s_1J_{\nu-1}(s_1)J_\nu(s_2)-s_2J_{\nu}(s_1)J_{\nu-1}(s_2)}{s_1+s_2}ds_1ds_2.\nonumber
\end{eqnarray}
This is a drastic simplification of the problem compared to Eq.~\eref{3.2.9}.

\section{Skew-orthogonal polynomials}\label{sec4}

What are the skew-orthogonal polynomials which correspond to the Pfaffian~\eref{3.6}? In order to solve this problem we consider the two-point measure~\eref{3.2}. The skew orthogonal polynomials $q_j$ are defined by
\begin{eqnarray}
 &&\int\limits_{\mathbb{R}^2}\det\left[\begin{array}{cc} q_{2j-1}(x_1) & q_{2j-1}(x_2) \\ q_{2i-1}(x_1) & q_{2i-1}(x_2) \end{array}\right]g(x_1,x_2)dx_1dx_2\nonumber\\
 &=& \int\limits_{\mathbb{R}^2}\det\left[\begin{array}{cc} q_{2j}(x_1) & q_{2j}(x_2) \\ q_{2i}(x_1) & q_{2i}(x_2) \end{array}\right]g(x_1,x_2)dx_1dx_2=0,\label{4.1}
\end{eqnarray}
and
\begin{eqnarray}
 \int\limits_{\mathbb{R}^2}\det\left[\begin{array}{cc} q_{2j+1}(x_1) & q_{2j+1}(x_2) \\ q_{2i}(x_1) & q_{2i}(x_2) \end{array}\right]g(x_1,x_2)dx_1dx_2=\widehat{h}_i^{(\nu)}\delta_{ij}.\label{4.2}
\end{eqnarray}
Moreover one has to assume that $q_l$ is a polynomial of order $l$.

The integral over the measure~\eref{3.2} for two arbitrary and conveniently integrable functions $f_1$ and $f_2$ can be simplified to
\begin{eqnarray}
 &&\int\limits_{\mathbb{R}^2}\det\left[\begin{array}{cc} f_1(x_1) & f_1(x_2) \\ f_2(x_1) & f_2(x_2) \end{array}\right]g(x_1,x_2)dx_1dx_2\nonumber\\
 &=&\frac{1}{2}\int\limits_0^\infty \det\left[\begin{array}{cc} f_1(x) & f_1(-x) \\ f_2(x) & f_2(-x) \end{array}\right]x^{2\nu}\exp\left[-nV(x^2)\right]dx.\label{4.3}
\end{eqnarray}
Due to this identity the skew-orthogonal polynomials $q_l^{(\nu)}$ are related by the orthogonal polynomials $p_l$ in the following way
\begin{equation}
 q_{2l}^{(\nu)}(x)=p_l^{(\nu)}(x^2)\label{4.4}
\end{equation}
for the even polynomials and
\begin{equation}
 q_{2l+1}^{(\nu)}(x)=xp_l^{(\nu)}(x^2)+{\rm const.}\,p_l^{(\nu)}(x^2)\label{4.5}
\end{equation}
for the odd polynomials. Notice that these skew-orthogonal polynomials for $V(x)=x$ (the Laguerre ensemble) are similar to but not completely the same as the one for $\beta=1$ and $\beta=4$ shown in Ref.~\cite{Meh04,Gho09} for the Laguerre ensemble. The reason is the two point weight which is
\begin{eqnarray}
 g_{\rm chGOE}(x_1,x_2)&=&(x_1x_2)^{\nu}\exp\left[-\alpha(x_1^2+x_2^2)\right]\frac{x_1-x_2}{|x_1-x_2|},\label{4.5a}\\
 g_{\rm chGSE}(x_1,x_2)&=&(x_1x_2)^{2\nu+3/2}\exp\left[-\alpha(x_1^2+x_2^2)\right]\delta^\prime(x_1-x_2),\label{4.5b}
\end{eqnarray}
in comparison see  Eq.~\eref{3.2} for $\beta=2$. The labels ``chGOE'' and ``chGSE'' refer to the chiral Gaussian orthogonal ensemble ($\beta=1$) and to the chiral Gaussian symplectic ensemble ($\beta=4$), respectively. The sign function $(x_1-x_2)/|x_1-x_2|$ generate the modulus of the Vandermonde determinant for $\beta=1$. The distribution $\delta^\prime$ is the first derivative of the Dirac delta function and cancels with these terms of the Vandermonde determinant which are zero at the support of the Dirac delta functions. This generates Cramers degeneracy in the quaternion case ($\beta=4$).

The solution of Eqs.~\eref{4.1} and \eref{4.2} is not unique which is reflected by the arbitrary constant in the odd polynomials~\eref{4.5}. One can readily confirm that this choice of the polynomials solves the conditions~\eref{4.1} and \eref{4.2} by recognizing the symmetry $q_{j}(-x)=(-1)^jq_j(x)$ and the orthogonality relation~\eref{2.13} for $p_j$. The normalization constant is
\begin{eqnarray}
 \widehat{h}_i^{(\nu)}=h_i^{(\nu)}.\label{4.6}
\end{eqnarray}
This relation between orthogonal and skew-orthogonal polynomials seems so trivial because of the particular and simple structure of the two-point weight~\eref{3.2}.

\section{A few more ensembles with Dyson index $\beta=2$ and Pfaffians}\label{sec5}

The algebraic rearrangement for chiral unitary ensembles described in Sec.~\ref{sec3} can be extended to other random matrix ensembles which have a squared Vandermonde determinant in the joint probability density function. By the same trick as in Eq.~\eref{3.1} we write
\begin{equation}\label{5.1}
 \Delta_{N}^2(z)=(-1)^{N(N-1)/2}\frac{\Delta_{2N}(\sqrt{z},-\sqrt{z})}{2^N\sqrt{\det z}},
\end{equation}
where the variables $z=\diag(z_1,\ldots,z_N)$ might be complex. The square root is the positive one but this is without loss of generality since the right hand side of Eq.~\eref{5.1} comprises both roots. Again the determinant of $z$ will be put to the measure $d\mu$ for a single eigenvalue.

We consider an average over ratios of characteristic polynomials for random ensembles like GUE and CUE, i.e.
\begin{eqnarray}\label{5.2}
 \widetilde{Z}_{k_1/k_2}^{(N)}(\kappa)=\int\limits_{\mathbb{C}^{N}}\Delta_{N}^2(z)\prod\limits_{i=1}^{N}\frac{\prod\limits_{j=1}^{k_2}(z_i-\kappa_{j2})}{\prod\limits_{j=1}^{k_1}(z_i-\kappa_{j1})}d\mu(z_i),
\end{eqnarray}
where $d\mu$ is a measure on $\mathbb{C}$ and $\kappa$ is chosen such that the integrals exist.  Notice that there is no modulus of the Vandermonde determinant which is a necessary property of the following discussion. A modulus of the Vandermonde is an obstacle to map Eq.~\eref{5.2} to the general joint probability density corresponding to the Pfaffian, see Ref.~\cite{KieGuh09b}, which we have not managed yet. A modulus corresponds to the biorthogonal polynomials \cite{Bergere:2003ht} whereas the choice without the modulus corresponds to the orthogonal polynomials, only. Apart from the modulus of the Vandermonde it is exactly the correlation function discussed in Sec.~4.2 of Ref.~\cite{KieGuh09a}.

With the help of the derivation in Sec.~\ref{sec3} the integral~\eref{5.2} can be written as
\begin{eqnarray}\label{5.3}
 \fl\widetilde{Z}_{k_1/k_2}^{(N)}(\kappa)&\propto&\frac{1}{\B_{k_1/k_2}(\sqrt{\kappa})}\\
 \fl&\times&\Pf\left[\begin{array}{c|c} \displaystyle\underset{}{\frac{\widetilde{d}(\sqrt{\kappa_{b2}}-\sqrt{\kappa_{a2}})}{\widetilde{Z}_{0/0}^{(\widetilde{d})}}\widetilde{Z}_{0/2}^{(\widetilde{d}-1)}(\kappa_{a2},\kappa_{b2})} & \displaystyle\frac{1}{\widetilde{Z}_{0/0}^{(\widetilde{d})}}\frac{\widetilde{Z}_{1/1}^{(\widetilde{d})}(\kappa_{b1},\kappa_{a2})}{(\sqrt{\kappa_{a2}}-\sqrt{\kappa_{b1}})} \\ \hline \displaystyle\frac{1}{\widetilde{Z}_{0/0}^{(\widetilde{d})}}\overset{}{\frac{\widetilde{Z}_{1/1}^{(\widetilde{d})}(\kappa_{a1},\kappa_{b2})}{(\sqrt{\kappa_{a1}}-\sqrt{\kappa_{b2}})}} & \displaystyle\frac{\sqrt{\kappa_{a1}}-\sqrt{\kappa_{b1}}}{(\widetilde{d}+1)\widetilde{Z}_{0/0}^{(\widetilde{d})}}\widetilde{Z}_{2/0}^{(\widetilde{d}+1)}(\kappa_{a1},\kappa_{b1}) \end{array}\right]\nonumber
\end{eqnarray}
for $k_1+k_2$ even and
\begin{eqnarray}\label{5.4}
 \fl&&\widetilde{Z}_{k_1/k_2}^{(N)}(\kappa)\propto\frac{1}{\B_{k_1/k_2}(\sqrt{\kappa})}\\
 \fl&\times&\Pf\left[\begin{array}{c|c|c} 0 & \displaystyle\underset{}{-\frac{\widetilde{d}}{\widetilde{Z}_{0/0}^{(\widetilde{d})}}\widetilde{Z}_{0/1}^{(\widetilde{d}-1)}(\kappa_{b2})} & \displaystyle\frac{1}{\widetilde{Z}_{0/0}^{(\widetilde{d})}}\widetilde{Z}_{1/0}^{(\widetilde{d})}(\kappa_{b1}) \\ \hline \displaystyle\frac{\widetilde{d}}{\widetilde{Z}_{0/0}^{(\widetilde{d})}}\widetilde{Z}_{0/1}^{(\widetilde{d}-1)}(\kappa_{a2}) & \displaystyle\underset{}{\frac{\widetilde{d}(\sqrt{\kappa_{b2}}-\sqrt{\kappa_{a2}})}{\widetilde{Z}_{0/0}^{(\widetilde{d})}}\widetilde{Z}_{0/2}^{(\widetilde{d}-1)}(\kappa_{a2},\kappa_{b2})} & \displaystyle\frac{1}{\widetilde{Z}_{0/0}^{(\widetilde{d})}}\frac{\widetilde{Z}_{1/1}^{(\widetilde{d})}(\kappa_{b1},\kappa_{a2})}{(\sqrt{\kappa_{a2}}-\sqrt{\kappa_{b1}})} \\ \hline \displaystyle-\frac{1}{\widetilde{Z}_{0/0}^{(\widetilde{d})}}\widetilde{Z}_{1/0}^{(\widetilde{d})}(\kappa_{a1}) & \displaystyle\frac{1}{\widetilde{Z}_{0/0}^{(\widetilde{d})}}\overset{}{\frac{\widetilde{Z}_{1/1}^{(\widetilde{d})}(\kappa_{a1},\kappa_{b2})}{(\sqrt{\kappa_{a1}}-\sqrt{\kappa_{b2}})}} & \displaystyle\frac{\sqrt{\kappa_{a1}}-\sqrt{\kappa_{b1}}}{(\widetilde{d}+1)\widetilde{Z}_{0/0}^{(\widetilde{d})}}\widetilde{Z}_{2/0}^{(\widetilde{d}+1)}(\kappa_{a1},\kappa_{b1}) \end{array}\right]\nonumber
\end{eqnarray}
for $k_1+k_2$ odd. The variable $\widetilde{d}$ is
\begin{equation}\label{5.5}
  \widetilde{d}=\left\{\begin{array}{cl} N+(k_2-k_1)/2, & k_2+k_1\in2\mathbb{N}, \\ N+(k_2-k_1+1)/2, & k_2+k_1+1\in2\mathbb{N}.\end{array}\right.
\end{equation}
The indices $a$ and $b$ run from $1$ to $k_1$ for $\kappa_1$ and from $1$ to $k_2$ for $\kappa_2$. Apart from the square roots of the variables $\kappa$ these structures are exactly the same as those of random matrix ensembles with Dyson index $\beta\in\{1,4\}$. Hence, it seems to be that the Pfaffian determinants~\eref{5.3} and \eref{5.4} for the average over characteristic polynomials are more general than the determinant derived in Ref.~\cite{KieGuh09a}.

Random matrix ensembles whose generating functions can be cast into the form~\eref{5.2} have this non-trivial expression as a Pfaffian. The Hermitian Gaussian unitary ensemble as well as its generalization with other potentials fulfill {\it a priori} this requirement since the joint probability density has a squared Vandermonde determinant without the modulus. More generally our derivation applies to each ensemble with a real spectrum, a squared Vandermonde determinant and a factorizing probability distribution, cf. Eq.~\eref{2.3}. Also the CUE (unitary group) can be cast into the form~\eref{5.2}. More ensembles can be found in the tables~1 and 2 of Ref.~\cite{KieGuh09a}. The Ginibre ensemble as well as its chiral counterpart are not in this class. Their joint probability density incorporates a modulus of the Vandermonde determinant and is, thus, in the class for the bi-orthogonal polynomials. Therefore it is possible that their eigenvalue correlation functions cannot be expressed in Pfaffians like Eq.~\eref{5.3} and \eref{5.4}.

The skew-orthogonal polynomials corresponding to the Pfaffians~\eref{5.3} and \eref{5.4} have the same relation to the orthogonal polynomials as chiral unitary ensembles, see Eqs.~(\ref{4.4}-\ref{4.6}). By construction this relation is so simple.

\section{Remarks and conclusions}\label{sec6}

We derived a non-trivial Pfaffian determinant for the average over ratios of characteristic polynomials of a large class of random matrix ensembles with Dyson index $\beta=2$. This structure is similar to the one for $\beta\in\{1,4\}$, cf. Ref.~\cite{KieGuh09b}. Hence, it is universal and unifies most of the symmetry classes known in the literature, particularly the Cartan classification \cite{Altland:1997zz,Zir96}. It is unclear how far beyond this classification \cite{Mag} this structure is applicable. It is only known that there are some of them which share the identity~\eref{3.6}. For example the real and quaternion Ginibre ensembles as well as there chiral counterpart fulfill an identity similar to Eq.~\eref{3.6}.

For many random matrix ensembles like the GUE it seems an academical question if one can derive a Pfaffian or not since there are no applications, yet. However, for the chiral GUE it is important to know this due to the new results obtained for the Wilson Dirac random matrix ensemble discussed in Refs.~\cite{Damgaard:2010cz,Akemann:2010zp,arXiv:1105.6229,arXiv:1108.3035,Kieburg:2011uf,KVZ11,VerSpl11}. Pfaffians were found there for finite lattice spacing. On the level of the joint probability density the authors of Ref.~\cite{arXiv:1108.3035} checked that the ensemble is the chiral GUE as well as the GUE at certain values of the lattice spacing. However for the eigenvalue correlation functions the continuum limit has not yielded the known determinant~\eref{2.20}. With this work we clarified this puzzle.

For intermediate ensembles in general our result might be helpful to understand the structure appearing by switching the interaction between the two ensembles on. It is numerically advantageous to think about spectral correlations of intermediate ensembles as kernels of Pfaffians since the integrand drastically simplifies. In combination with the supersymmetry method \cite{Guh06,Som07,LSZ08,KGG09} also the number of integrals reduces a lot.

Moreover, we derived the relation between the orthogonal polynomials and the skew-orthogonal polynomials corresponding to the determinants and the Pfaffians, respectively. This relation, see Eqs.~(\ref{4.4}-\ref{4.6}), is not only quite simple but also universal since it applies to all random matrix ensembles discussed in this work. The relation between the orthogonal and skew-orthogonal polynomials for $\beta=2$ slightly differs to those found in Ref.~\cite{Gho09} for the cases $\beta=1$ and $\beta=4$. The difference in the two-point weight is the reason for this. Based on the representations~\eref{2.11}, \eref{5.3} and \eref{5.4} shared by all random matrix ensembles with $\beta=2$ as well as checks of these representations \cite{Ver11}, we conjecture that the recursion relation of the orthogonal polynomials connects the determinant and the Pfaffian and this has to be done in a general way.

The Pfaffian found for the average over characteristic polynomials carries over to the $k$-point correlation functions. This structure is valid in the large matrix limit, too. It should not depend on which scaling limit is chosen since the Pfaffian is independent of the matrix size. Hence, the correlation functions appearing as kernels of this Pfaffian have non-trivial recursion relations mapping the determinant to the Pfaffian.

\section*{Acknowledgements}

I am grateful to Gernot Akemann, Jacobus J.M. Verbaarschot and Savvas Zafeiropoulos for fruitful discussions and helpful comments. I also thank Peter J. Forrester and Christopher D. Sinclair for pointing out their work~\cite{Sin,ForSin}. Furthermore I acknowledge financial support by the Alexander-von-Humboldt Foundation.

\end{document}